\documentclass[twocolumn,amsmath,amssymb,aps, prl,
showkeys,showpacs, superscriptaddress, floatfix] {revtex4-2}
\usepackage{graphicx}
\usepackage[usenames,dvipsnames]{color}
\usepackage{bm}
\usepackage[colorlinks=true,breaklinks=true,allcolors=blue]{hyperref}
\usepackage{dcolumn}
\usepackage{slashed}
\usepackage{mathtools}
\usepackage{multirow}

\def\ri{{\rm i}}
\def\re{{\rm e}}
\allowdisplaybreaks

\begin{document}
\title{Topological Interfaces of Luttinger Liquids}
\author{Ananda Roy}
\email{ananda.roy@physics.rutgers.edu}
\affiliation{Department of Physics and Astronomy, Rutgers University, Piscataway, NJ 08854-8019 USA}
\author{Hubert Saleur}
\email{saleur@usc.edu}
\affiliation{Institut de Physique Th\'eorique, Paris Saclay University, CEA, CNRS, F-91191 Gif-sur-Yvette}
\affiliation{Department of Physics and Astronomy, University of Southern California, Los Angeles, CA 90089-0484, USA}

\begin{abstract}
Topological interfaces of two-dimensional conformal field theories contain information about symmetries of the theory and exhibit striking spectral and entanglement characteristics. While lattice realizations of these interfaces have been proposed for unitary minimal models, the same has remained elusive for the paradigmatic Luttinger liquid {\it i.e.,} the free, compact boson model. Here, we show that a topological interface of two Luttinger liquids can be realized by coupling special one-dimensional superconductors. The gapless excitations in the latter carry charges that are specific integer multiples of the charge of Cooper-pairs. The aforementioned integers are determined by the windings in the target space of the bosonic fields -- a crucial element required to give rise to nontrivial topological interfaces. The latter occur due to the perfect transmission of certain number of Cooper-pairs across the interface. The topological interfaces arise naturally in Josephson junction arrays with the simplest case being realized by an array of experimentally-demonstrated~$0-\pi$ qubits, capacitors and ordinary Josephson junctions. Signatures of the topological interface are obtained through entanglement entropy computations. In particular, the subleading contribution to the so-called interface entropy is shown to differ from existing field theory predictions. The proposed lattice model provides an experimentally-realizable alternative to spin and anyon chains for the analysis of several conjectured conformal fixed points which have so far eluded ab-initio investigation.
\end{abstract}
\maketitle 
Boundaries and interfaces in two-dimensional conformal field theories~(CFTs) contain signatures of their universal characteristics~\cite{Cardy1989, Petkova:2000ip}. Investigation of the associated finite-size effects serves as a powerful method of characterizing the corresponding critical system. The latter are relevant for a large family of physical problems ranging from impurity scattering in condensed matter physics~\cite{Affleck1995conformal, Saleur1998, Saleur2000} to Dirichlet branes in string theory~\cite{Polchinski1995, Douglas:1999vm}. 

Topological or perfectly-transmissive interfaces~\cite{Petkova:2000ip, Frohlich2004} are those that maintain the continuity of the stress-energy tensor across the interface. Thus, they can be deformed without affecting the values of the correlation functions as long as they are not taken across field insertions. The defect operators associated with the topological interfaces reflect the internal symmetries of the CFT~\cite{Frohlich2006} and play a fundamental role in the investigation of anyon chains~\cite{Feiguin:2006ydp}, generalized notions of symmetries~\cite{Cordova:2022ruw, Seiberg:2023cdc} and the correspondence between two-dimensional CFTs and three-dimensional topological field theories~\cite{Fuchs:2002cm, Buican2017}. 

Given the importance of topological interfaces in CFTs, it is natural to construct lattice models that realize them in the scaling limit, thereby enabling ab-initio computations of the interface characteristics. This is particularly important for the investigation of properties that are not directly amenable to analytical computations without making additional assumptions. An example is the ground state entanglement entropy~(EE) in the presence of topological interfaces, where violations of the field theory predictions due to the existence of zero energy modes have been found for the Ising~\cite{Roy2021a, Rogerson:2022yim} and the Potts~\cite{Sinha:2023hum} models. In fact, lattice realizations of arbitrary topological interfaces in unitary minimal models are known based on  quantum spin or anyon chains~\cite{Grimm1992, Grimm2001, Aasen2016, Belletete2023, Sinha:2023hum}. However, the same has remained elusive for Luttinger liquids {\it i.e.,} free, compactified boson models~\footnote{For realization of non-topological interfaces using coupled XXZ spin chains, see Refs.~\cite{Venuti2009, Roy2021b}}. This is due to the difficulty of systematic realization of integer~($>1$) winding of the bosonic field in target space using spin/anyon chains. 
\begin{figure}
\centering
\includegraphics[width = 0.5\textwidth]{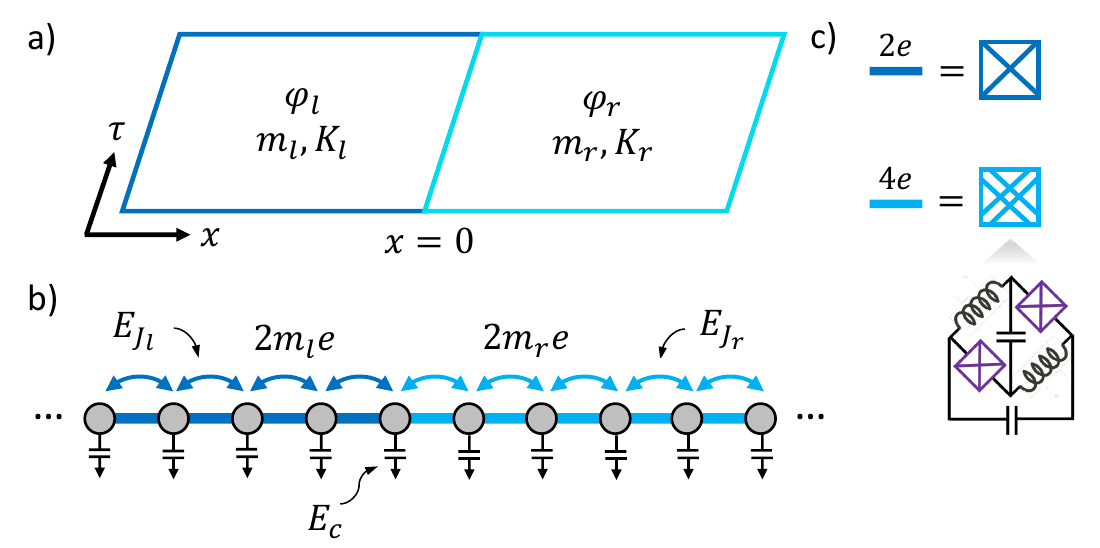}
\caption{\label{fig:schematic} a) The bosonic fields~$\varphi_l, \varphi_r$ are `glued' at the interface located at~$x=0$. The Luttinger parameters~(winding numbers in target space) on either side of the interface are denoted by~$K_l, K_r$~($m_l, m_r$). The topological interface occurs for~$m_l^2K_r = m_r^2K_l$. b) One-dimensional Josephson-junction array that realizes the interface of panel a) in the scaling limit. The superconducting granules~(gray circles), with charging energy to ground plane~$E_c$, are separated by generalized Josephson junctions. The latter allow tunneling of only~$m_{l(r)}$ Cooper-pairs with charge~$2m_{l(r)}e$ on the left~(right) of the interface with rate~$E_{J_{l(r)}}$. c) Schematics of an ordinary Josephson junction and a~$0-\pi$ qubit~\cite{Doucot2002, Ioffe2002, Kitaev2006c, Brooks2013} required to realize the simplest nontrivial topological interface with~$m_l = 1, m_r = 2$.}
\end{figure}

Here, we show that arbitrary integer windings in target space and thus, nontrivial topological interfaces of free boson models, can be realized by coupling one-dimensional superconductors with gapless excitations whose charges are specific integer multiples of the charge of the Cooper-pairs. The latter Hamiltonians arise naturally in Josephson junction arrays that allow tunneling of only the aforementioned specific integer multiples of Cooper-pairs. The topological interface occurs for certain ratios of the Luttinger parameters that govern the decay of vertex operators on either side of the interface. The relevant continuum model and the lattice realization are described below. 

Fig.~\ref{fig:schematic}(a) shows the schematic with the interface located at~$x=0$ for the free, compact, bosonic fields~$\varphi_l, \varphi_r$ with Luttinger parameters~$K_l, K_r$. At the interface, the fields satisfy~$m_r\varphi_l(x= 0, \tau) = m_l\varphi_r(x=0,\tau)$, where the integers~$m_{l,r}$ are the corresponding winding numbers~\cite{Bachas2001}. The topological interface is a special case arising when~$m_l^2K_r = m_r^2K_l$~\cite{Bachas2007}. Since~$m_l = m_r = 1$ leads to~$K_l = K_r$ {\it i.e.}, no interface, nontrivial cases arise only when at least one of the winding numbers is greater than 1. For the sake of definiteness, we consider~$m_{l}, m_{r}$ coprime with~$m_l<m_r$. 

Fig.~\ref{fig:schematic}(b) shows the schematic of the lattice model that realizes the aforementioned interface in the scaling limit. The gray circles denote superconducting islands with charging energy~$E_c$ arising due to the self-capacitance. On the left~(right) of the interface, the superconducting islands are separated by generalized Josephson junctions that allow tunneling of only~$m_l(m_r)$ Cooper-pairs at a rate~$E_{J_{l(r)}}$. The case~$m_{l,r} = 1$ corresponds to the ordinary Josephson junction, while~$m_{l,r} = 2$ corresponds to the recently-proposed~\cite{Doucot2002, Ioffe2002, Kitaev2006c, Brooks2013} and experimentally-demonstrated~\cite{Gladchenko2008, Smith2020, Gyenis2021}~$0-\pi$ qubit~[Fig.~\ref{fig:schematic}(c)]. The Hamiltonian for the lattice model with open boundary condition is given as
\begin{align}\label{eq:H_latt}
H &= E_c\sum_{j=1}^Ln_j^2 - E_{J_l}\sum_{j = 1}^{j_0-1}\cos\left(m_l\phi_j - m_l\phi_{j+1}\right)\nonumber\\&\qquad- E_{J_r}\sum_{j = j_0}^{L-1}\cos\left(m_r\phi_j - m_r\phi_{j+1}\right).
\end{align}
Here, $n_j$ is the excess number of Cooper pairs on the~$j^{\rm th}$ superconducting island and $\phi_k$ is the superconducting phase at the~$k^{\rm th}$ node, satisfying~$[n_j, {\rm e}^{\pm \ri\phi_k}] = \pm\hbar\delta_{jk}{\rm e}^{\pm \ri\phi_k}$~\footnote{We consider the case when the energy scale~$E_{C_J}$ associated with the junction capacitance of each of the generalized Josephson junctions is much smaller than~$E_c, E_{J_l}$, $E_{J_r}$ so that phase-slips across the generalized junctions are exponentially suppressed. }. For~$E_{J_l}, E_{J_r}\gg E_c$, the portion of the array on the left~(right) of the interface is in a superconducting state with gapless excitations carrying charge~$2pem_{l(r)}$,~$p\in\mathbb{Z}$~\footnote{This is a direct generalization of the case considered in Ref.~\cite{Bradley1984}. The model with only onsite repulsion is chosen for the sake of simplicity. We checked that the topological fixed point analyzed in this work persists in the presence of nearest-neighboring repulsion~\cite{Glazman1997}.}. The low-energy properties are described by the Euclidean action
\begin{align}
\label{eq:S}
S &= \frac{1}{2\pi K_l}\int_{x<0}\left(\partial_\mu\varphi_l\right)^2 + \frac{1}{2\pi K_r}\int_{x>0}\left(\partial_\mu\varphi_r\right)^2
\end{align}
with the interface condition~$m_r\varphi_l(x=0) = m_l\varphi_r(x=0)$. 
The bosonic fields on either side of the interface are compactified with radius~$2\pi$:~$\varphi_{l,r} \equiv \varphi_{l,r} + 2\pi$. The vertex operators~$\re^{\ri\varphi_{l,r}}$ correspond to the coarse-grained counterpart of the lattice operators~$\re^{\ri m_{l,r}\phi_j}$. The correlation-functions of $\re^{\ri p\varphi_{l(r)}}$,~$p\in\mathbb{Z}$ on the left~(right) side of the array decay algebraically with an exponent determined by~$K_{l(r)}$. The latter depend on the ratios~$E_{J_{l,r}}/E_c$~\cite{Roy2020a}. The condition for the interface to be topological,~$m_l^2K_r = m_r^2K_l$, ensures that the vertex operators~$\re^{\ri m_r\varphi_l}$ and~$\re^{\ri m_l\varphi_r}$ decay with the same exponent on the two sides of the interface. Note that the Hamiltonian~$H$~[Eq.~\eqref{eq:H_latt}] directly provides a lattice regularization of the fixed point action~[Eq.~\eqref{eq:S}]. The same interface~(topological or otherwise) could also be realized by replacing the first term in the final summation in Eq.~\eqref{eq:H_latt} by ~$-\cos[m_lm_r(\phi_{j_0} - \phi_{j_0 + 1})]$. This would lead to a contribution to the continuum action of the form~$-M\int d\tau \cos[m_r\varphi_l(0) - m_l\varphi_r(0)]$ where~$M>0$ is a coupling constant. From dimensional analysis, as long as~$m_l^2K_r + m_r^2K_l<2$, the aforementioned interface condition~[below Eq.~\eqref{eq:S}] would be realized in the infrared limit.

\begin{figure}
\centering
\includegraphics[width = 0.48\textwidth]{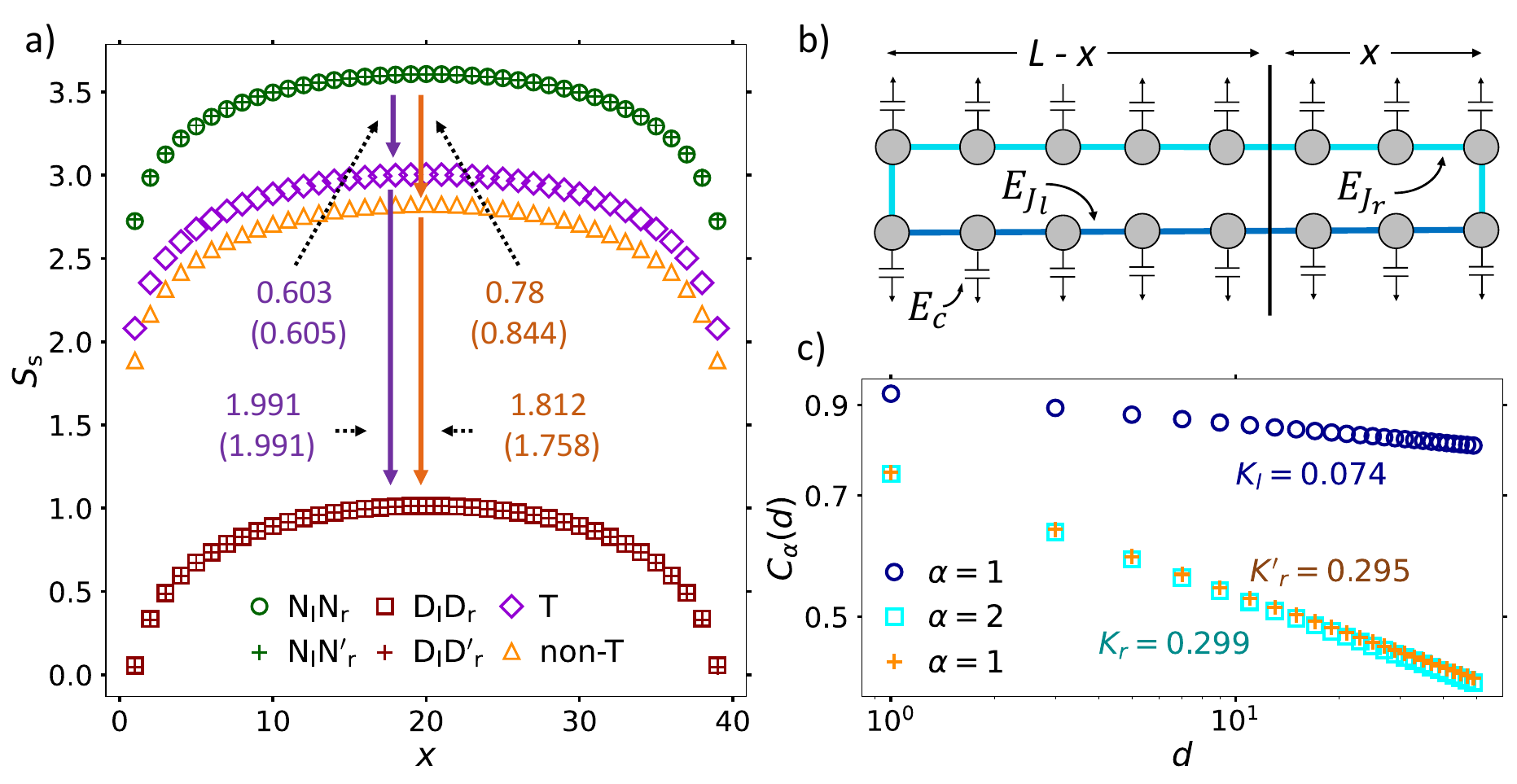}
\caption{\label{fig:fig_2} DMRG results in the folded configuration with identical boundary condition on both ends. The length of the folded array,~$L = 40$ and the winding numbers are chosen as:~$m_l = 1, m_r = 2$. (a) EE~$(S_{\rm s})$ as a function of subsystem size for varying boundary conditions. Here, $N_lN_r$ and~$D_lD_r$ and T correspond to Neumann~(free), Dirichlet~(fixed) and topological boundary conditions on both ends of the folded array.  In particular,~$N_lN_r$~($D_lD_r$) correspond to free~(fixed) boundary condition on both~$\varphi_l$ and~$\varphi_r$ with boundary g-function~$g_b = g_{N_l}g_{N_r}$~$(g_{D_l}g_{D_r})$. The parameters~$E_{J_l},E_{J_r}$ are chosen such that the corresponding Luttinger parameters on the two halves of the array are~$K_r \approx 4K_l$. To emphasize the role of winding numbers,~$S_{\rm s}$ is also computed for another case:~$m_l = m'_r = 1$ with~$K'_r \approx 4K_l$. The corresponding free~$(N_lN^{'}_r)$, fixed~$(D_lD^{'}_r)$ and non-T~(non-topological) boundary conditions are shown. The EE  for the $N_lN^{'}_r (D_lD^{'}_r)$ case coincides with that for $N_lN_r~(D_lD_r)$~[Eq.~\eqref{eq:g_fun}]. However, the EE is {\it not} the same for the non-T and T boundary conditions, reflected in the difference between the orange and violet curves. This is due to a difference in the corresponding g-functions.~(b) Schematic of the configuration analyzed.~(c) The Luttinger parameters~$K_l, K_r$ and~$K'_r$ are obtained using infinite DMRG computations of correlation functions:~$C_\alpha(d) = \langle \re^{\ri\alpha\phi_{j}}\re^{-\ri\alpha\phi_{j+d}}\rangle$,~$\alpha = 1,2$.}
\end{figure}

Next, numerical simulation results, obtained with the density matrix renormalization group~(DMRG) technique~\cite{DMRG_TeNPy}, are presented for the proposed lattice model. The diagnostic of choice is the g-function or the universal `ground state degeneracy'~\cite{Affleck1991} associated with the interface. Here, the logarithm of the g-function is obtained by computing the subleading~${\cal O}(1)$ term in EE at a conformal quantum critical point~\cite{Calabrese2004, Calabrese2009}. In the context of interfaces, this subleading term arises in the EE of a region symmetrically located around the interface. Folding the system at the interface  identifies the g-function as that associated with a boundary condition for a CFT with central charge twice that of the unfolded model. In the case where the boundary conditions at the ends of the folded model are the same~[Fig.~\ref{fig:fig_2}(b)], the EE for the subsystem of size~$x$ is
\begin{equation}
\label{eq:S_s}
S_{\rm s}(x) = \frac{c}{3}\ln\left[\frac{2L}{\pi}\sin\frac{\pi x}{L}\right]+ \ln \left(g_b g_{N_l} g_{N_r}\right) + \ldots,
\end{equation}
where ~$c = 1$, the central charge of the gapless bulk theory and the dots indicate non-universal contributions. Among the three universal contributions to the~${\cal O}(1)$ term,~$\ln g_b$ arises due to the boundary condition of the folded array, while the remaining two due to the Neumann boundary conditions at the entanglement cuts in the lower and upper halves of the folded array~\cite{Cardy2016, Roy2021b}. For the interface of the two bosonic fields realized in Eqs~(\ref{eq:H_latt}, \ref{eq:S}),~$g_b = g_{\rm int}$. The relevant g-functions are given as~\cite{Bachas2007}
\begin{equation}
\label{eq:g_fun}
g_{N_{\alpha}} = K_\alpha^{-\frac{1}{4}}, g_{D_{\alpha}} = \frac{K_\alpha^{\frac{1}{4}}}{\sqrt{2}}, g_{\rm int} = \left[\frac{m_l^2K_r + m_r^2K_l}{2\sqrt{K_lK_r}}\right]^{\frac{1}{2}}
\end{equation}
where~$\alpha = l,r$~\footnote{The formulas in Ref.~\cite{Bachas2007} can be recovered using the definition:~$R_\alpha = 1/\sqrt{K_\alpha}$,~$\alpha = l,r$. We adhere to the definition of the topological interface based on the Luttinger parameter since it is the more physically relevant quantity in the superconducting setting.}. At the topological point, the interface g-function depends only on the winding numbers and equals~$\sqrt{m_lm_r}$. 

\begin{figure}
\centering
\includegraphics[width = 0.48\textwidth]{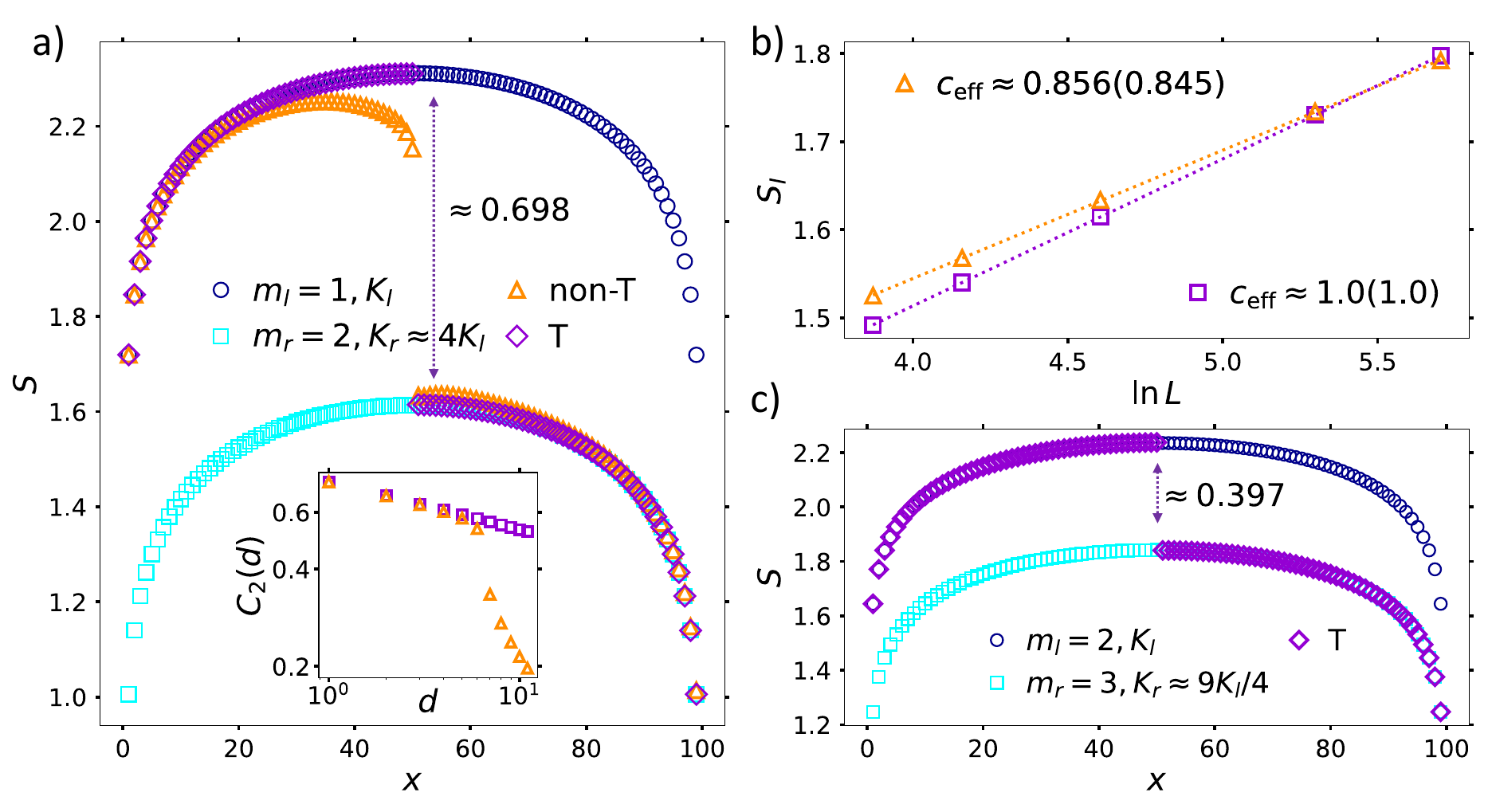}
\caption{\label{fig:fig_3} DMRG results for the ground-state EE for different bipartitionings of the array in Fig.~\ref{fig:schematic}(b) with open boundary conditions at the ends. The parameters~$K_l, K_r$ and~$K'_r$ are chosen as in Fig.~\ref{fig:fig_2}. a) The blue~(cyan) circles~(squares) correspond to the EE for the ground state if only the left~(right) free boson theory occupied the entire array. The violet~(orange) curve shows the case of the topological~(non-topological) interface, denoted by T~(non-T). For the topological~(non-topological) interface, the Luttinger parameters and winding numbers for the bosonic field on the right are chosen to be~$K_r\approx 4K_l$($K^{'}_r\approx 4K_l$) and~$m_r = 2~(m'_r = 1)$ respectively. Only in the case of $m_r = 2$ is the topological interface realized. This is manifest in the EE seamlessly turning in to that for the bosonic field on the right as well as in the correlation functions~$C_2(d) = \langle\re^{2\ri\phi_j}\re^{-2\ri\phi_{j+d}}\rangle$~(inset). The latter exhibit a continuous~(abruptly-changing) behavior across the interface for the topological~(non-topological) case.~(b) The topological~(non-topological) nature of the interface for the violet~(orange) curves of panel a) is verified by the variation of the interface EE with the logarithm of the subsystem-size~(expected values in the parenthesis).~(c) EE as a function of subsystem size for the interface between bosonic fields with winding numbers~$m_l = 2, m_r = 3$. The parameters~$E_{J_l}, E_{J_r}$ are chosen such that~$K_l\approx0.086, K_r\approx0.196\approx9K_l/4$. In both panels a) and c), the change in the EE at the interface~$\approx\ln(m_r/m_l)$, which is different from earlier predictions~\cite{Sakai2008}.}
\end{figure}

Fig.~\ref{fig:fig_2}(a) shows the EE results for the folded chain for several choices of boundary conditions. The latter are chosen to be identical for the two ends. The topological interface is analyzed for~$m_l = 1, m_r = 2$. The parameters~$E_{J_l}, E_{J_r}$ are chosen such that the Luttinger parameters satisfy~$K_r \approx 4K_l$. The green~(maroon) circles~(squares) correspond to Neumann~(Dirichlet) boundary conditions for two bosonic fields. The Neumann case is realized by keeping the two halves of the array decoupled, while the Dirichlet boundary condition by applying a strong boundary potential of the form~$-\cos[m_{l(r)}\phi_k]$ for the lower~(upper) halves of the array~\cite{Roy2020a}. Here, the index~$k$ denotes the site-indices at the ends of the folded array. Finally, results for the topological boundary condition~(purple diamonds) are obtained by analyzing the ground state of~$H$~[Eq.~\eqref{eq:H_latt}]. The change in the universal contribution to the EE is obtained by computing the change in the EEs at the center of the chain,~$x\approx L/2$~\footnote{This method is more robust than fitting to Eq.~\eqref{eq:S_s} and relies on the fact that the size of the system is much larger than the length-scale associated with the boundary perturbation.}, with the expected results being shown in parentheses. The relevant Luttinger parameters are obtained from the exponent of the algebraic decay of the correlation functions~$C_\alpha(d) = \langle \re^{\ri\alpha\phi_{j}}\re^{-\ri\alpha\phi_{j+d}}\rangle$,~$\alpha = 1,2$ using infinite DMRG~[Fig.~\ref{fig:fig_2}(c)]. Next, it is checked that the topological interface arises truly as a consequence of the winding numbers and not just due to the Luttinger parameters having a certain ratio. To that end, results are also shown for the case when the winding numbers for the bosonic fields on the two sides of the interface are:~$m_l = m^{'}_r = 1$, with the Luttinger parameters~$K'_r \approx 4K_l$. In this case, when Neumann and Dirichlet boundary conditions are imposed for the two bosons at both ends, the EE remains the same. This is because~$g_{N_\alpha}, g_{D_\alpha}$ depend only on the Luttinger parameters and not on the winding numbers. However, when the upper and lower arrays are coupled, the EE {\it is} different from the topological case~[Eq.~\eqref{eq:g_fun}], as is also confirmed by the numerical simulations~(orange curve labelled non-top)~\footnote{The discrepancy with the expected result is due to finite-size effect. That this is the case was verified by computing different system-sizes. Note that the corrections to scaling for the non-topological case appear to be larger than the topological case. This interesting effect deserves to be investigated further.}.

Next, the perfectly transmissive nature of the topological interface is verified from the behavior of the interface EE. The latter is the EE of a subsystem chosen with its one end coinciding with the interface and the other with the end of the system~\cite{Sakai2008}. The interface EE depends logarithmically on the system-size~$S_I = (c_{\rm eff}\ln L)/6 + S_0$. For the model under consideration, the coefficient~$c_{\rm eff}$ defines an effective central charge which depends continuously on the interface coupling~\cite{Sakai2008, Eisler2010} and equals the central charge of the bulk theory only for a topological interface. The universal part of the subleading term~$S_0$, when computed using the twisted torus partition function, has been shown to depend on the winding numbers of the bosonic fields glued at the interface~\cite{Sakai2008}. However, this analytical computation is known to be problematic since it fails to be capture the true geometric configuration of the interface. The corresponding prediction for the Ising model~\cite{Brehm2015} has already been shown to be incorrect using ab-initio computations~\cite{Roy2021a, Rogerson:2022yim}. The free boson case is no different, as shown below.

Fig.~\ref{fig:fig_3}(a) shows the DMRG results for the EE for different bipartitionings of the unfolded array with open boundary conditions at the ends~[Fig.~\ref{fig:schematic}(b)]. The blue~(cyan) circles~(squares) show the EE for the ground-state of the left~(right) bosonic theory. The purple diamonds show the EE for the topological~(T) case when the EE transitions smoothly from the EE of the left bosonic theory to that of the right. This smooth transition is symbolic of the perfectly-transmissive interface. The corresponding variation of the interface EE with system-size is shown in Fig.~\ref{fig:fig_3}(b) where the central charge is indeed obtained to be 1. On the other hand, the EE for the non-topological~(non-T) interface~(orange triangles, same parameters as Fig.~\ref{fig:fig_2}) shows a clear dip in the EE indicating the existence of reflections of the entanglement-carrying modes. The corresponding variation with subsystem-size in Fig.~\ref{fig:fig_3}(b) reveals a coefficient of the leading logarithmic term~$\approx0.856$. This is close to~$0.845$, obtained using the analytical predictions of Ref.~\cite{Sakai2008}. However, the change in the subleading term {\it is different} from what Ref.~\cite{Sakai2008} predicts for both T and non-T cases. This subleading term measured with respect to the EE of the left bosonic theory, for a topological interface, is simply given by the difference between the EEs of the two bosonic models. This difference equals~$2\ln \left(g_{N_r}/g_{N_l}\right)$. The logarithms of the g-functions arise due to the Neumann boundary conditions at the entanglement cuts for the two bosonic models~\cite{Cardy2016, Roy2021b}. At the topological point, this leads to an offset~$\ln(m_r/m_l)$, which is {\it different} from $-\ln (m_lm_r)$ obtained in Ref.~\cite{Sakai2008}. To verify this, the EE for different bipartitionings is shown for the topological interface realized with winding numbers~$m_l=2, m_r = 3$. Note that the subleading term in the interface EE assumes this simple form only for the topological interface. Analytical prediction away from the topological point remains unknown to us. Finally, the inset of the panel a) shows the behavior of the pair-correlation function~$C_2(d)=\langle \re^{2\ri\phi_{j}}\re^{-2\ri\phi_{j+d}}\rangle$. The latter across a topological interface~(violet diamonds), exhibits a continuous behavior as expected from perfect transmission of the pairs of Cooper-pairs. However, across the non-topological interface~(orange triangles), the slope changes abruptly at the interface, which is expected since the same correlation function decays four times faster on the right of the interface. 

To summarize, this work investigates Josephson-junction arrays that realize topological interfaces of Luttinger liquids. These interfaces, requiring specific integer windings in target space of the bosonic fields, are those that allow perfect transmission of the same integer number of Cooper-pairs across the interface. The signatures of the topological interface are obtained through EE computations of the ground state. The defect g-function, obtained by considering the change in the boundary entropy in the folded picture, is in agreement the analytical results for the~$c = 2$ CFT. On the other hand, the change in the subleading term in the scaling of the interface EE is found to be~$\ln(m_r/m_l)$, which is different from the existing analytical predictions~\cite{Sakai2008}. 

The Josephson-junction based incarnation of the topological interface is amenable to experiments, with the simplest nontrivial case requiring only experimentally-demonstrated quantum circuit elements. Signatures of perfect transmission of integer number of Cooper-pairs, reminiscent of Andreev reflection~\cite{Andreev1965}, as well as potential formation of bound-states for a `CFT bubble'~\cite{Bachas2001} could be measured in transport experiments. The tunable Josephson junction allows investigation of observables both at and away from the topological fixed point. Unlike the boundary CFT for the free, compact boson that has been analyzed extensively in the context of quantum brownian motion~\cite{Caldeira1983}, superconductor-insulator transition~\cite{Schmid1983, Lukyanov:2007xu, Murani2020, Kuzmin2023} and tunneling in a one-dimensional electronic systems~\cite{Kane1992, Fendley1995a}, the same for the~$c = 2$ CFT remains much less explored. Several boundary fixed points have been predicted to occur~\cite{Furusaki1993, Wong1994, Lesage1998, Yi1998, Affleck2001} with exotic properties that do not have counterparts for the~$c = 1$ theory. The proposed lattice model and its generalizations provide a systematic way to realize several of the aforementioned fixed points that have so far eluded lattice investigation. 

Discussions with Pasquale Calabrese, Sergei Lukyanov and Erik Tonni are gratefully acknowledged. AR was supported from a grant from the Simons Foundation (825876, TDN). HS was supported by the French Agence Nationale de la Recherche (ANR) under grant ANR-21-CE40-0003 (project CONFICA).
\bibliography{/Users/ananda/Dropbox/Bibliography/library_1}
\end{document}